\documentclass[11pt,twoside]{article}

\usepackage{asp2004}
\usepackage{epsf}
\usepackage{psfig}
\usepackage{lscape}

\begin{document}
\title{Dust Production in the High-Redshift  Universe}   
\author{S. V. Marchenko}   
\affil{Department of Physics and Astronomy, Western Kentucky University,
Bowling Green, KY 42101-3576, USA}    

\begin{abstract} 
How much dust can be produced in the early Universe?
Does dust production depend on the average heavy-metal
content of the hosting galaxy? Considering supernova explosions,  massive
stars (Wolf-Rayet, LBV and RSG), and relatively massive AGB
stars among possible dust-generating objects in the early Universe, we
find that SN remnants can be regarded as the main source
of the primordial dust. However, this conclusion is based on
highly uncertain (and probably over-estimated) dust production rates.
Despite all the uncertainties, interstellar extinction must be taken into
account while observing high-redshift objects.
\end{abstract}


\section{Introduction}

Contemplating the general importance of dust, one should take into account the following:
$\sim 50\%$ of optical radiation emitted since the Big Bang by all astrophysical sources
has been `reprocessed' by dust. Dust is regarded as an efficient ISM coolant \citep{leh04}, thus
controlling accretion and profoundly influencing star-formation rates, especially
for massive stars \citep[e.g.][]{wol87}. Assembly of $H_2$ on dust grains is far more
efficient than in the gas phase. On cosmological scales, dust may distort the cosmic
microwave background and
change the far-IR background \citep{elf04}. There are first indications of the presence of dust in
high-redshift Lyman-break galaxies \citep{and04,ouc04} and quasars \citep{mai04}, which may have serious implication
for the estimated star-formation rates \citep[e.g.][]{sch05}.

\section{Dust Formation in the Milky Way Galaxy  and in the High-Redshift Universe}

The detailed galactic census of dust-producing stars \citep{geh89} proves that in the modern
Universe, relatively low-mass stars dominate the scene.
All categories of evolved, post-main-sequence objects (Miras, carbon stars,
late-type supergiants, etc.) account for $\sim 90\%$ of the stellar
dust output, while planetary nebulae produce $<1\%$. Their high-mass counterparts,
supernovae and Wolf-Rayet stars, amount to $<10\%$, combined. Dust grains are
also manufactured via relatively slow-paced accretion in molecular clouds: $(1-5)\times$ the
stellar output. Apparently, in the 1-2 Gyr-old Universe only the high-mass, $M\ga 3M_\odot$, stellar
population may be responsible for accumulation of copious amounts of dust (Table 1).
In the Table 1 we provide: (a) the main categories of dust-producing stars, (b) their initial masses,
(c) the total dust yield, per star, (d) the chemistry and (e) the size distribution of dust grains,
where `st.' (standard) corresponds to the truncated power-law distribution \citep{mat77}, and
`sm.' (small) is used to emphasize the presence of particles with $a\ll 1\mu$ sizes. We separate the
cases of normal, solar metallicity and low-Z environments.

\begin{table}[!ht]
\caption{Potential dust-generating stars in the $z>3$ Universe}
\smallskip
\begin{center}
{\scriptsize
\begin{tabular}{llcccl}
\tableline
\noalign{\smallskip}
Category & $M_{ini}$ &  & \multicolumn{1}{c} {Dust} & & Ref. \\
\noalign{\smallskip}
\cline{3-5}
\noalign{\smallskip}
       &  $M_\odot$ & mass/*  ($M_\odot$) & composition & size & \\
\noalign{\smallskip}
\tableline
\noalign{\smallskip}
SN$_{\gamma\gamma}$(PopIII) &140-260 & ?              ,$Z\ll Z_\odot$        & ? & ? & \\
\noalign{\smallskip}
SN II                                            & $\ge8$  & 0.1-0.3, $Z\ll Z_\odot$            & {\tiny Si,C,Fe,Al,Mg,O}  & `st.' (?) & 1 \\
                                                    &               & $\la$1, $Z\sim Z_\odot$        & {\tiny Si,C,Fe,Al,Mg,O}  & `st.', $\sim 1\mu$ &  2,3  \\
\noalign{\smallskip}
LBV                                         &$\ge75-85$&    ?          ,$Z\ll Z_\odot$         & ? & ? &     \\
                                                   &$>30(?)$ & 0.01-0.25, $Z\sim Z_\odot$  & {\tiny Si,C,PAH,Al(?)}  & $\sim 1\mu$+`sm.' & 4,5,6,7   \\
\noalign{\smallskip}
WCd                                       &$\ge60-70$&    ?          ,$Z\ll Z_\odot$         & ? & ? &    \\
                                                   &$\ge30 $ & {\tiny $10^{-3}-10^{-2}$,}$Z\sim Z_\odot$  & {\tiny C(amorph.)}  & $\sim 1\mu$+`sm.' & 8,9,10   \\
\noalign{\smallskip}
sgB[e]                                     &       ?          &   ?          ,$Z\ll Z_\odot$         & ? & ? &    \\
                                               &$\ge30-60 $&  ? ,$Z\sim Z_\odot$  & Si  & $\sim 1\mu$ &   11  \\
\noalign{\smallskip}
B[e]WD                                   &    $\ge5$   &   ?                                         & ? & ? &      \\
\noalign{\smallskip}
RSG                                        & $\ge8-50 $&   ?         ,$Z\ll Z_\odot$         & ? & ? &    \\
                                               &$\ge8-25 $ &  {\tiny $10^{-4}-10^{-3}$,}$Z\sim Z_\odot$  & Si  & $\sim 0.5\mu$+`st.' &  12,13,14  \\
\noalign{\smallskip}
AGB                                        & $\ge2-(5-6)$  & {\tiny $\propto M_{ini}(Z/Z_\odot)$},$Z\ll Z_\odot$         & ? & ? &      \\
(OH/IR)                                   &$\ge2-6 $ &  {\tiny $\approx10^{-3}M_{ini}$,}$Z\sim Z_\odot$  & Si,C,ice  & `st.' (?) &   \\
\noalign{\smallskip}
\noalign{\smallskip}
\tableline
\end{tabular}
}
\end{center}

[1] \citet{mai04}, [2] \citet{cla01}, [3] \citet{sug05}, [4] \citet{wat97}, [5] \citet{voo99}, [6] \citet{voo00}, [7] \citet{kin02},
[8] \citet{wil87}, [9] \citet{mat92}, [10] \citet{mar02}, [11] \citet{mol02}, [12] \citet{sea89}, [13] \citet{jur96},
[14] \citet{smi01}\\

\end{table}

Judging by the multitude of question marks, not much is known about dust formation
in low-Z environments. The pair-instability supernovae \citep[SN$_{\gamma\gamma}$: ][]{heg02} belong to
the broad category of Population III objects. Though their general characteristics and ability to produce dust
are yet to be established, one may assume their dust yields to be comparable with
the present-epoch SN events, thus making them the major dust sources in the high-redshift
universe. However, one should notice the substantial difference between the total dust outputs
of SN events provided by
different research groups: the relatively high, $M_{dust}\sim 1M_\odot$, estimates of \citet{dun03} and
\citet{mor03}, in line with theoretical expectations \citep{tod01},
vs. the $M_{dust}\sim 10^{-3}M_\odot$ values from \citet{dwe92}, \citet{dwe04} and \citet{poz04}.
There is a strong indication that at least in some SN events \citep[the Crab nebula, SN2002hh and SN2002ic:][]{gre04,bar05,kot05} dust
comes from a progenitor, either a luminous blue variable (LBV), red supergiant (RSG) or carbon-rich Wolf-Rayet
star (WCd). With the lower yields for SN events, the integral output from RSG and WCd stars may rival
the SN category. The relatively rare B supergiants with forbidden emission lines, sgB[e] \citep{lam98}, as
well as newly discovered (and probably associated with sgB[e]) class of early-type, luminous stars with
warm circumstellar dust, B[e]WD \citep{mir05}, could be considered as minor contributors, unless the number
of related sources is grossly under-estimated. One more category emerges
at $z<9$ and gradually comes to a complete dominance over the dust production in the modern universe: the
asymptotic giant-branch stars (AGB). This group shows a clear dependence of dust production on Z
\citep{loo00}. This dependence is less pronounced for SNe \citep{tod01}. For the remaining categories of massive
dust-producing stars one may assume that dust production depends on the mass loss rate and use the
general \.M=f(Z) relationship of \citet{vin01}.

\section{Interstellar Extinction for High-Redshift Objects}

Estimating an average extinction for high-redshift objects, we adopt and slightly modify the approach of \citet{loe97}.
Namely, the dust absorption coefficient, $\alpha_\nu$, is expressed as
$$ \alpha_\nu(z,Z)=\rho_{dust} (z, Z)\kappa_\nu(Z),$$
with dust opacity $\kappa_\nu(Z)$ represented by the Galactic law \citep{mat90} for a $Z\sim Z_\odot$
environment, or the Small Magellanic Cloud dependence \citep{car05} for $Z< Z_\odot$; z defines the redshift
and Z denotes the metallicity. The dust density takes
the form
$$\rho_{dust} (z, Z)= \Omega_b \rho_c (1+z)^3 \sum _i F_i(z) f_{dep, i}(z) f_{dust,i}(Z),$$
where the sum runs over different categories of dust producers (Table 1),
$\rho_c=9.7\,10^{-30}g\, cm^{-3}$ provides the current critical density of the universe, and  $\Omega_b=0.044$
gives the total baryonic density. The mass fraction of dust, $f_{dust,i}(Z),$, deposited by a given star
depends on the
category of dust-producing stars and the ambient metallicity, while the mass fraction of stars being able
to produce dust is calculated as
$$f_{dep, i} (z)= \int \limits _{M1_i} ^ {M2_i} m^{-(1+x)} dm \, /\,  \int \limits _{M_d(z)} ^ {M_u(z)} m^{-(1+x)} dm,$$
with $M_d(z), M_u(z)$ and $x$ depending on the redshift, and  $M1_i, M2_i$ provided by Table 1. Hence, the variable $x$ could be anywhere
between x=0.5 (`top-heavy' mass function) and 1.35 (classical form) at $z\ge 10$, then  converging  to
x=1.35 for $z<10$. We define the $F_i(z)$ term as
$$F_i(z)=\int \limits _z ^{20} \eta_i (z') \frac{dF_{col}}{dz'}exp(-\frac{t_z-t_{z'}}{T})
f_{star}(z')dz',$$
where the efficiency of star formation $f_{star}(z)$ is lowered by the presence of compact objects (neutron stars,
black holes, white dwarfs) in the overall `recycling' loop: $f' _{star}(z)=f_{star}(z)(1-\xi f_{star}(z+dz))$, with
$\xi =0.10-0.15$ and $f_{star}(z)$ taken from \citet{dro05}. The upper limit, z=20, is imposed by the
evolution of Population III objects  \citep[e.g.][]{cho05}. We adopt the mass fraction of  baryons assembled into
collapsed objects, $F_{col}(z)$, following \citet{hai97}. The $exp(-(t_z-t_{z'})/T)$ term provides the dust
survival timescales, with T ranging from 0.1 to 1.0 Gyr \citep{dra79}. The term $\eta_i (z)$ introduces evolutionary timescales
(i.e. cutoffs) for different categories of dust producers: $\eta_i (z)=1 \mbox{ for $z\le z_{cr,i}$}$, and  $\eta_i (z)=0
\mbox{ for $z > z_{cr,i}$}$.
Then, the optical depth of a dusty medium at a given redshift $z_s\ge3$ is
$$\tau_{dust}(\nu , z_s ,Z) =\frac{c}{H_0}\int \limits _3 ^ {z_s} \frac{\alpha_{\nu (1+z)}(z,Z)}
{(1+z)^{5/2}} dz ,$$
with $H_0$ and c as universal constants. Following the arguments of \citet{loe97},
we ignore dust production at  $z<3$, as it will be generally confined to individual
galaxies (dominance of AGB stars; hence, rather low velocities of dust ejecta)
rather than $\sim$homogeneously distributed along the line of sight.

\begin{figure}[ht!]
\plotfiddle{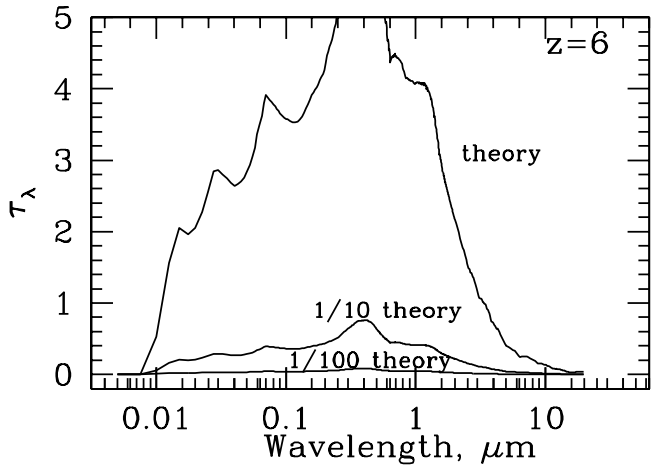}{2.0in}{0}{100}{100}{-300}{-300}
\caption{The dust opacity for a z=6 object with variable yield from the SN ejecta.}
\end{figure}

Exploring the different parameters of the model, we confirm the conclusion of \citet{loe97} that
dust chemistry has relatively small influence on the overall opacity.  More serious is the uncertainty
in the dust yields of SNe. Considering the theoretical predictions \citep[e.g.][]{tod01}, we find that
they result in an inappropriately high $\tau_{dust}$ for a z=6 object (Fig. 1). Lowering them by an
order of magnitude brings the theoretical yields closer to some estimates from recent observations
\citep[e.g.][]{bar05}.
With the appropriately adjusted SN output, the average dust opacity may be neglected for
$z\sim5$ objects, {\it unless they pass through a star-burst episode}, and should
be taken into account for $z\gg5$ objects (Fig. 2).

\begin{figure}[ht!]
\plotfiddle{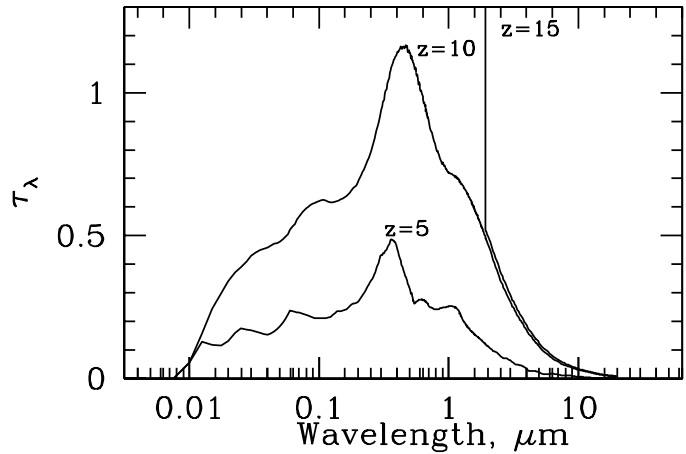}{1.9in}{0}{100}{100}{-300}{-300}
\caption{The average dust opacity for objects with variable z. We assume
that the re-ionization continues until z=15.}
\end{figure}

Grouping the dust-producing stars into 3 general categories: SN events vs. massive stars
(RGB, LBV, WCd) vs. AGB, and varying the yield from the SN stars, as well as adjusting the
dependence of the dust production rate on Z (metallicity) for the massive stars, we plot two
extreme scenarios in Fig. 3. Considering the shares of dust production in the modern
universe, one may expect that the model with the lowest SN yield and absence of a steep
dependence of the dust production rate in massive stars on the ambient metallicity  (right panel of Fig. 3) may be
closer to reality.

The calculations also show that, on average, the 'survival' timescales of the primordial
dust, $T\le 4\, 10^8$ yr,  closely match the current-epoch expectations for the Galaxy \citep{jon94}.

\begin{figure}[ht!]
\plotfiddle{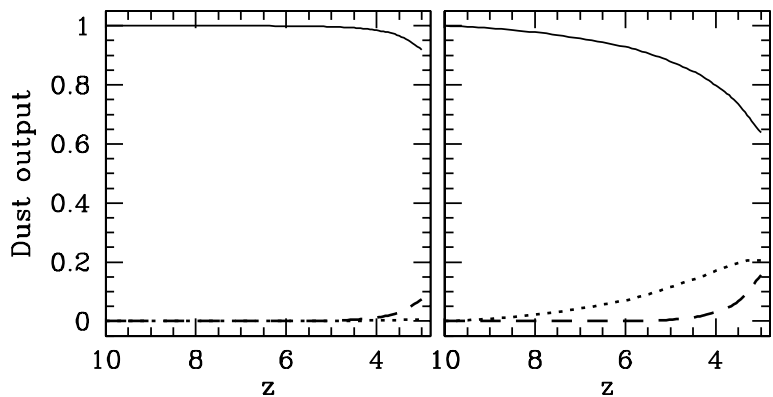}{1.75in}{0}{100}{100}{-220}{-306}
\caption{Three main groups of dust-producing stars: SN (full lines), massive stars (RSG+LBV+WCd: dotted
lines) and AGB (dashed lines). Left panel: the SN rate is lowered to 1/10 of the theoretical predictions and
the dust production rate for massive stars depends on Z. Right panel: SN rate =1/30 theoretical, no dependence
on Z for massive stars.}
\end{figure}

\section{Conclusions}

\begin{itemize}

\item SN events should be regarded as a major source of dust in the high-redshift ($z>3$) universe.
This conclusion may be independent of the source of dust, unless different sources provide
substantially different extinction curves \citep[e.g.][]{mai04}. It could be either primordial dust coming from
a progenitor (LBV, RSG or WCd star), or dust produced in the SN ejecta. However, the most important
issue is the ability of dust to survive in the hostile environment: shocks, UV radiation. On an optimistic note, one
may mention the case of WCd stars where, facing a similar challenge as in the shocked environments
of SN ejecta, the grains of amorphous carbon
manage to survive for {\it {at least}} $10^2$ years \citep{mar02}, thus effectively reaching (and enriching) the ISM.

\item Our calculations show that, in order to be comparable to the known output of SN events in the Galaxy
\citep{geh89}, the theoretical estimates of dust production in SNe populating the 1-2 Gyr-old universe
should be lowered by an order of magnitude, thus providing a better match to the recent \citep[e.g.][]{bar05}
observations.

\item Overall, IS extinction should be appropriately taken into account for all $z\gg5$ objects in order to
estimate their true properties, especially realizing that for some of them, due to the enhanced star-formation
rate, the local extinction may substantially exceed the average values (cf. Fig. 2).

\end{itemize}

\acknowledgements S.M. is thankful for the financial help provided by the Office of Sponsored
Programs and the Dean's Office, WKU.


\end{document}